\documentstyle[multicol,aps,epsfig]{revtex}

\tighten       

\begin{document}
\draft 

\title{Effects of disorder on the wave front depinning
transition in spatially discrete systems}
\author{A. Carpio$^1$, L. L. Bonilla$^2$ and A. Luz\'on$^3$}
\address{$^1$Departamento de Matem\'{a}tica
Aplicada, Universidad Complutense de Madrid,\\
28040 Madrid, Spain\\
$^2$Departamento de Matem\'{a}ticas, Escuela
Polit\'ecnica Superior,  Universidad Carlos III
de Madrid,\\ 
Avenida de la Universidad 30, 28911 Legan{\'e}s,
Spain \\
$^3$Escuela T\'ecnica Superior de Ingenieros de Montes, 
Universidad Polit\'ecnica de Madrid,\\ 
28040 Madrid, Spain}
\date{ 7 December 2001  }
\maketitle

\begin{abstract}
Pinning and depinning of wave fronts are ubiquitous
features of spatially discrete systems describing
a host of phenomena in physics, biology, etc. A
large class of discrete systems is described by
overdamped chains of nonlinear oscillators with
nearest-neighbor coupling and subject to random external
forces. The presence of weak randomness shrinks the pinning
interval and it changes the critical exponent of the wave
front depinning transition from 1/2 to 3/2. This effect is
derived by means of a recent asymptotic theory of the
depinning transition, extended to discrete drift-diffusion
models of transport in semiconductor superlattices and
confirmed by numerical calculations. 
\end{abstract}
\pacs{05.45.-a; 82.40.Bj; 45.05.+x}

\begin{multicols}{2}
\narrowtext
Phenomena in many different fields may be described by means
of spatially discrete systems: motion of dislocations
in crystals \cite{nab67}, atoms adsorbed on a periodic
substrate \cite{cha95}, arrays of coupled diode resonators
\cite{diode}, weakly coupled semiconductor superlattices (SL)
\cite{bon94,car01}, sliding of charge density waves (CDW)
\cite{cdw}, superconductor Josephson array junctions
\cite{jj}, propagation of nerve impulses along myelinated
fibers \cite{kee87,kee98}, pulse propagation through cardiac
cells \cite{kee98}, calcium release waves in living cells
\cite{bug97}, etc. In many of these systems, disorder due to
differences in the parameters of individual elements is
important because it has a strong impact in the collective
behavior. A distinctive example of collective behavior in
discrete systems (not shared by continuous ones) is the
phenomenon of wave front pinning: for values of a control
parameter in a certain interval, wave fronts joining two
different constant states fail to propagate \cite{kee98}.
When the control parameter surpasses a threshold, the wave
front depins and starts moving \cite{kee87}. The existence
of such thresholds is an intrinsically discrete fact, which
is lost in continuum aproximations. Recently, a theory of
front depinning and motion near threshold has been proposed
by some of us for one-dimensional (1D) nonlinear spatially
discrete reaction-diffusion (RD) systems \cite{car01PRL}. In
our theory, propagation failure and front depinning are
characterized by studying the behavior of a few sites,
provided the effects of spatial discretization are
sufficiently strong \cite{car01,car01PRL}. 

In this paper, we consider the effect of weak disorder on the
wave front depinning transition in spatially discrete 1D
systems. Applications will include discrete RD systems
subject to a random field, sliding CDW and domain motion in
SL. In these examples, the main effect of disorder is to
soften the transition, changing the critical exponent from
1/2 to 3/2. The latter value was obtained by D.\ Fisher in a
mean field model of sliding CDW using  scaling arguments
\cite{fis83}.

We consider chains of diffusively coupled overdamped
oscillators in a  potential $V$, subject to a random force
field $F + \gamma\xi_n$:
\begin{equation}
{du_{n}\over dt} = u_{n+1}-2u_n + u_{n-1} + F - A\,  g(u_n)
+ \gamma\xi_n.   \label{Fd}
\end{equation}
Here $g(u)=V'(u)$ presents a `cubic' nonlinearity, such that
$A\, g(u)-F$ has three zeros, $U_1(F/A) <U_2(F/A) <U_3(F/A)$
in a certain force interval ($g'(U_i(F/A)) >0$ for $i=1,3$,
$g'(U_2(F/A))<0$). The fluctuating part of the force field
is $\gamma\xi_n$, where $\gamma\geq 0$ characterizes the
disorder strength and $\xi_n$ is a zero mean random variable
taking values on an interval $(-1,1)$ with equal
probability. An example of a model described by Eq.\
(\ref{Fd}) is (except for the mean field approximation which
we do not make) D.\ Fisher's modification of the
Fukuyama-Lee model of sliding CDW \cite{fis83}. In it, $u_n=
\theta_n - \chi_n$, $g(u)=\sin u$, $\gamma\xi_n= \chi_{n+1}
-2\chi_n +\chi_{n-1}$, where $\theta_n$ is the CDW phase at
the site $n$ and $\chi_n$ is a random variable taking values
with equal probability on $(0,2\pi)$. 

Provided $g(u)$ is odd with respect to $U_2(0)$ and $\gamma=
0$, there is a symmetric interval $|F|\leq F_c$ where the
wave fronts joining the stable zeros $U_1(F/A)$ and
$U_3(F/A)$ are pinned. For $|F|>F_c$, there are {\em smooth
traveling wave fronts}, $u_n(t)= u(n-ct)$, with $u(-\infty)=
U_1$ and $u(\infty) =U_3$. The velocity $c(A,F)$ depends on
$A$ and $F$ and it satisfies $cF<0$ and $|c|\propto (|F|-
F_c )^{ {1\over 2}}$ as $|F|\to F_c$ \cite{car01PRL}.
Examples are the overdamped Frenkel-Kontorova (FK) model
($g=\sin u$) \cite{FK} and the quartic double well potential
($V=(u^2-1)^2 /4$). Less symmetric nonlinearities yield a
non-symmetric pinning interval and our analysis of the
depinning transition applies to them with trivial
modifications \cite{car01}. 

Let us recall the main features of the active point theory
of the wave front depinning transition in the absence of
disorder \cite{car01PRL,car01}. Except when $A$ is too small
(the continuum limit in which $F_c\to 0$), the stable wave
front profile differs appreciably from either $U_1$ or $U_3$
at finitely many points, $u_n$, $n=-L, \ldots, -1,0,1,
\ldots, M$, called the {\em active points}. At $n<-L$,
$u_n\approx U_1(F/A)$ and at $n>M$, $u_n\approx U_3(F/A)$.
We shall reconstruct the wave front profile $u(n-ct)$ by
analyzing the behavior of the active points $u_n(t)$, $n=-L,
\ldots, -1,0,1,\ldots, M$, as the front moves for $F>F_c>0$
(the case $F<0$ can be obtained by using symmetry
considerations). The arbitrary phase of the (translation
invariant) wave front will be fixed by imposing that the
solution of the system of active points at time $t=0$ (and
$F$ slightly larger than $F_c$) be equal to its stationary
solution at $F=F_c$, $u_n(A,F_c)$, $n=-L, \ldots, M$ up to
terms of order $(F-F_c)$. $F_c$ is obtained from the
condition that the matrix of the coefficients in the system
of active points linearized about the stationary solution
has a zero eigenvalue. Provided $V_n$ ($V_{-L}^2+\ldots+
V_M^2 =1$) is the corresponding eigenvector, an outer
approximation to the solution $u_n(t)$ is $u_n(t) \sim
u_n(A,F_c) + \varphi(t) V_n$, where the amplitude $\varphi$
obeys the equation $d\varphi/dt = \alpha (F-F_c) + \beta
\varphi^2$, in which $\alpha=\sum_{i=-L}^M V_i + A^{-1}\,
[V_{-L}/ g'(U_1(F_c/A)) + V_{M}/ g'(U_3(F_c/A))] >0$,
$\beta= -(A/2) \sum_{i=-L}^M g''(u_i(A,F_c)) V_i^3 >0$. The
solution of this equation such that $\varphi(0)=0$
[equivalent to $u_n(0) = u_n(A,F_c)$ for $n=-L,\ldots, M$]
is $\varphi = [\alpha (F-F_c)/\beta]^{{1\over 2}}
\tan\big(\sqrt{\alpha \beta (F-F_c)}t\big)$. The amplitude
$\varphi$ blows up at times $t=\pm t_b$, $t_b=\pi /(2\sqrt{
\alpha\beta (F-F_c)})$. The inverse width of this time
interval yields an approximation for the wave front
velocity, $|c|\sim \sqrt{\alpha \beta (F-F_c)}/\pi$. At the
blow up times, the previous outer approximation to $u_n(t)$
has to be matched to an appropriate inner solution. At the
later blow up time $t_b$, the appropriate inner solution is
the solution of the active point system at $F=F_c$ with the
boundary conditions that $u_n= u_n(A,F_c)$ as $t\to -\infty$
and $u_n= u_{n+1}(A, F_c)$ as $t\to\infty$. At the earlier
blow up time $-t_b$, the inner solution obeys the same
system of equations at $F=F_c$, but the boundary conditions
are $u_n= u_{n-1}(A, F_c)$ as $t\to - \infty$ and $u_n=
u_{n}(A, F_c)$ as $t\to \infty$ \cite{carSIAP}. 

{\em Effects of  disorder}. How does weak disorder modify
this picture of the wave front depinning transition? Our
main idea is to find a dominant balance of the disorder
effects with nonlinearities and $(F-F_c)$ near the depinning
transition. Given our active point construction sketched
above, the dominant balance is struck provided $(F-F_c) =
O(\gamma)$ as $\gamma\to 0$. The amplitude Equation becomes 
\begin{eqnarray}
{d\varphi\over dt} = \alpha (F-F_c) + \gamma \sum_{n=-L}^M
V_n\xi_n + \beta \varphi^2 , \label{ameq}
\end{eqnarray}
and the matching condition is the same as before. The
solution of Eq.\ (\ref{ameq}) blows up at the end of time
intervals of duration $1/|c_R|$, where
\begin{eqnarray}
|c_{R}| = {1\over\pi}\,\sqrt{\alpha \beta (F-F_c) + \gamma
\beta \sum_{n=-L}^M V_{R+n}\xi_{R+n}} ,  \label{cn}
\end{eqnarray} 
provided the argument of the square root is positive.
Otherwise the motion of the wave front stops and it becomes
pinned. Notice that we have chosen now $u_R(t)$ as the
central active point that was distinguished with the
subscript zero in our previous formulas. After the blow up,
$u_R$ jumps to $u_{R+1}(A,F_c)$, approximately, and it
remains there until a time $1/|c_{R+1}|$ has elapsed. Then
it jumps to $u_{R+2}(A,F_c)$ approximately, and so on. 

The magnitude of interest in these systems is usually an
average velocity $|c_R|$ over sufficiently many points. For
example, this magnitude is proportional to the current due to
a sliding CDW and it is important to know its behavior near
the depinning field and the magnitude thereof. We shall argue
that the average velocity is approximately given by the
following equation:
\begin{eqnarray}
|\overline{c}| \equiv {1\over N} \sum_{R=1}^N |c_{R}| =
\langle |c_R(\xi)|\rangle,   \label{c1}\\
\langle |c_{R}(\xi)|\rangle \equiv {1\over 2\pi}
\int_{-1}^{1} \{\alpha \beta (F-F_c) + \gamma \beta 
\sigma\xi\}_+^{{1\over 2}}\, d\xi.   \label{c2}
\end{eqnarray} 
Here $N\gg (L+M+1)$ is sufficiently large, $\sigma=1$, and
$\{x\}_+^{{1\over 2}}$ is $\sqrt{x}$ if $x>0$ and zero
otherwise. The idea of the proof is as follows. Let us
assume that $A$ is so large that $L=M=0$ and there is only
one active point. Then the central limit theorem applied to
Eq.\ (\ref{cn}) with $V_R=1$ yields Eqs.\ (\ref{c1}) -
(\ref{c2}). Let us assume now that there are two active
points. The previous argument fails because now the sum in
Eq.\ (\ref{cn}) comprises two terms instead of one. Then the
terms in the arithmetic mean are no longer independent: when
$u_R=u_1$ for instance, Eq.\ (\ref{cn}) contains $\xi_1$ and
$\xi_2$. After the blow up time, $u_R=u_2$ and Eq.\
(\ref{cn}) contains $\xi_2$ and $\xi_3$, etc. However, we
can group the sums appearing in the arithmetic average of
Eq.\ (\ref{c1}) in two groups containing only independent
random variables: $R=2r-1$ and $R=2r$, with $r=1, 2,\ldots$.
The variable $V_1\xi_1 + V_2\xi_2$ has zero mean and
correlation $2\sigma^2/3$, where $\sigma^2= V_1^2 + V_2^2
=1$. This correlation is exactly the same as that of the
variable $\xi_1$. Then the central limit theorem applied to
each group (of sums of `dimer' random variables) gives one
half the integral in Eq.\ (\ref{c2}), and the sum of these
two halves yields Eq.\ (\ref{c1}). If we have more active
points, we just have to subdivide the arithmetic mean in as
many subgroups as active points and use the previous
argument to prove Eq.\ (\ref{c1}). 

\begin{figure}
\begin{center}
\includegraphics[width=8cm]{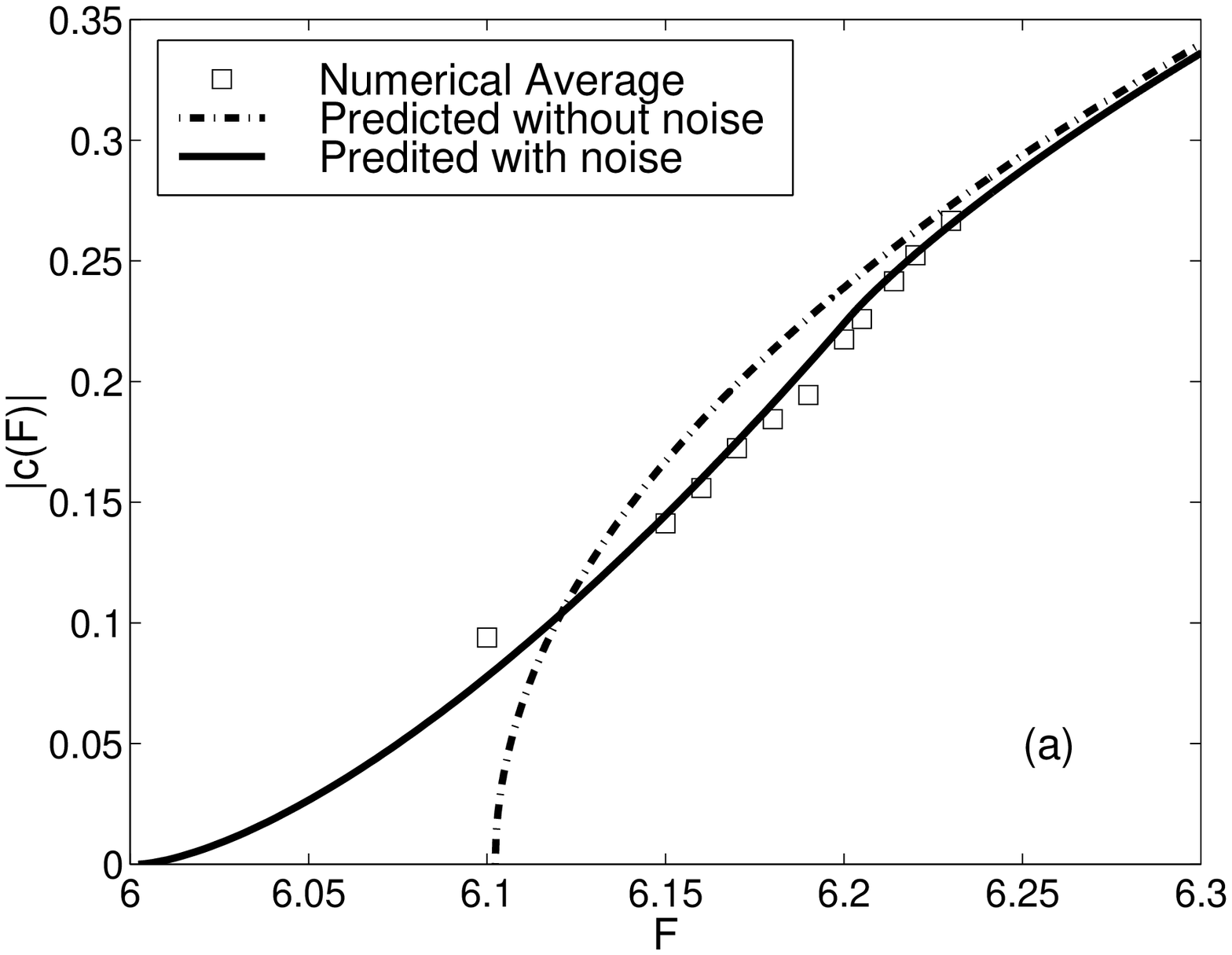}
\includegraphics[width=8cm]{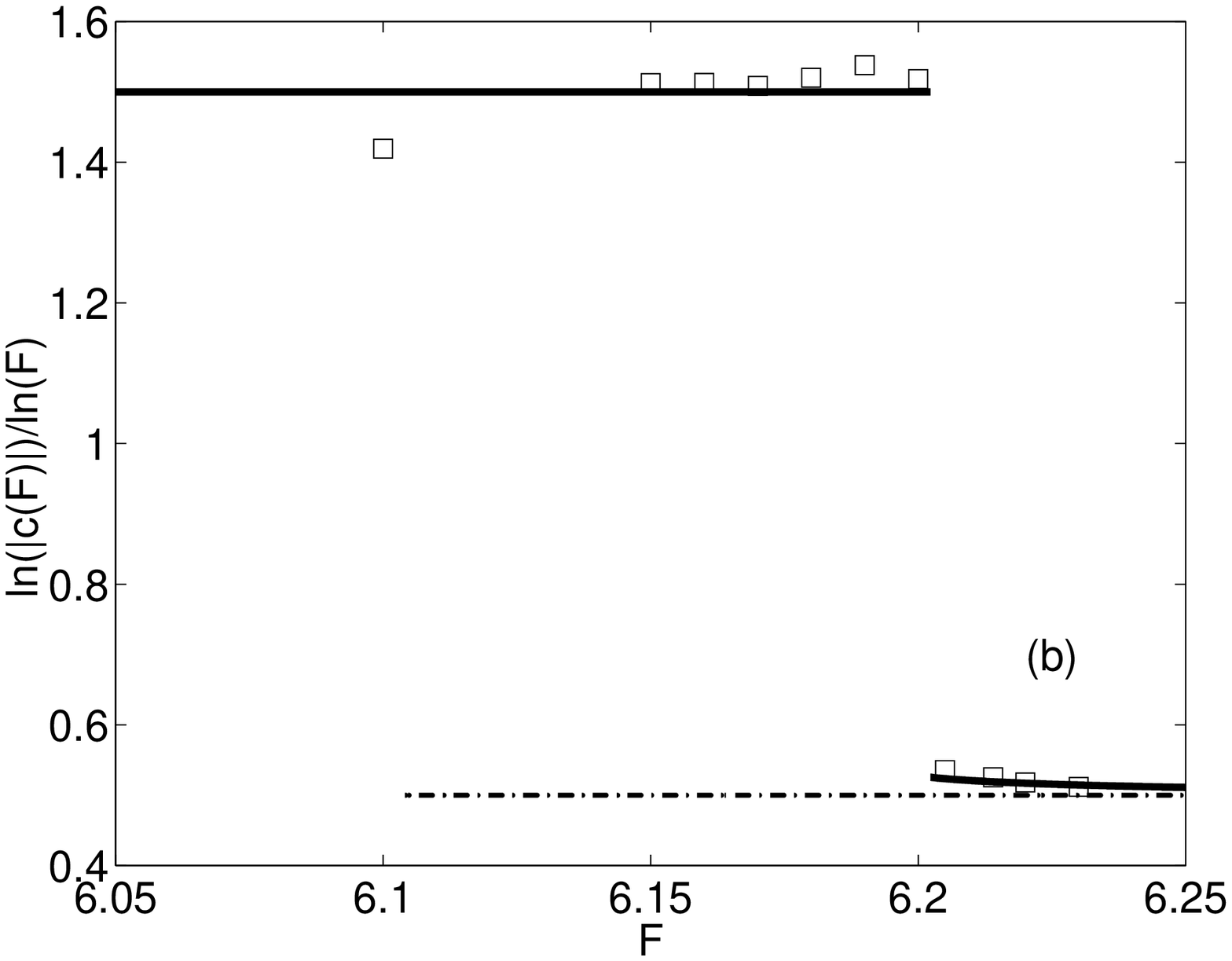}
\caption{(a) Average velocity $|\overline{c}|$ as a function
of $F$ for $A=10$, $F_c = 6.102281$ and $\gamma=0.1$. (b)
Graph of ln$|\overline{c}|$/ln$F$ showing the crossover to
the critical exponent $3/2$. }
\label{fig1}
\end{center}
\end{figure}

The elementary integral in Eq.\ (\ref{c2}) yields
$\overline{c} =0$ if $F<F_c -\gamma\sigma/\alpha$ (recall
that $\sigma=1$),
\begin{equation}
|\overline{c}| = {\sqrt{\beta\sigma}\over 3\pi\gamma}
\left\{
\begin{array}{l} |{\alpha\over\sigma} (F-F_c)+\gamma|^{{3
\over 2}},\,\, \mbox{if}\,\, |F-F_c|\leq {\gamma\sigma\over
\alpha}, \\  
|{\alpha\over\sigma} (F-F_c) +\gamma|^{{3\over 2}} - 
|{\alpha\over\sigma} (F-F_c) -\gamma|^{{3\over 2}} 
\end{array}\right.  \label{c3}
\end{equation} 
if $(F-F_c)>\gamma\sigma/\alpha$. Clearly we have a new
critical field $F_c^* = F_c -\gamma\sigma/\alpha$, and a new 
critical exponent $3/2$ (instead of $1/2$ for the case
without disorder), $|\overline{c}| \propto (F-F_c^*)^{3/2}$.
We have compared our theory to the direct numerical solution
of Eq.\ (\ref{Fd}) in Fig.~\ref{fig1}, obtaining an
excellent agreement of theoretical predictions and numerical
simulation. 

\begin{figure}
\begin{center}
\includegraphics[width=8cm]{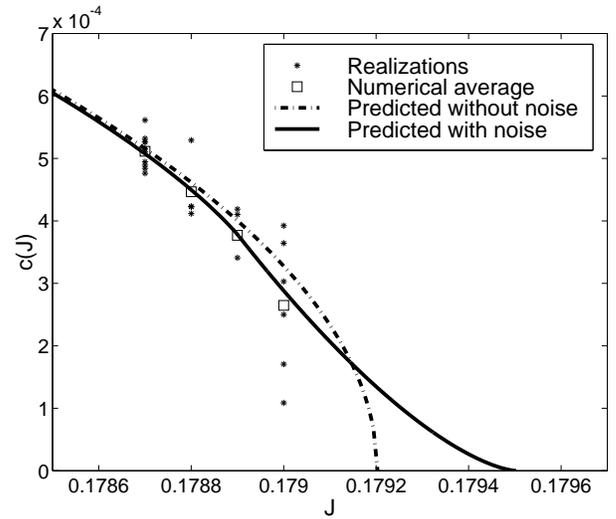}
\caption{ Dimensionless average wave front velocity for the
9/4 SL with dimensionless parameters $\nu=3$, $J_1=
0.179203$, $\gamma= 0.01$. }
\label{fig2}
\end{center}
\end{figure}

\begin{figure}
\includegraphics[width=8cm]{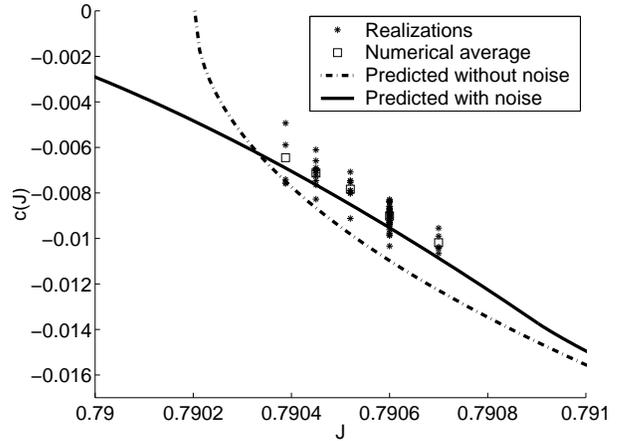}
\begin{center}
\caption{Same as Fig.~\ref{fig2} with parameters $\nu= 40$,
$J_2= 0.790203$, $\gamma= 0.01$.}
\label{fig3}
\end{center}
\end{figure}

Similarly, we can analyze the effect of weak disorder in
the doping of the wells on the motion of wave fronts in {\em
dc} current biased semiconductor SL. If the total current
density is close to a pinning value, the displacement
current is almost zero except at certain times during wave
front motion at which the wave front jumps from one well to
the adjacent one. The average velocity given by Eq.\
(\ref{c1}) is proportional to the arithmetic mean of time
averages of the displacement current over the time interval
between maxima thereof. The average velocity is basically
the mean velocity at which the wave front traverses $N$
wells. To calculate it, we take advantage of our theory of
wave front motion in current biased SL \cite{car01}. The
equations we use are those in Ref.~\onlinecite{car01} except
that the dimensionless Poisson equation is now $E_i -
E_{i-1} = \nu (n_i -1 -Ê\gamma\xi_i)$ where $\gamma$ and
$\xi$ defined as in Eq.\ (\ref{Fd}) represent the disorder
in well doping. In the SL equations, the roles of the force
$F$ and the parameter $A$ are taken by the total current
density $J$ and the dimensionless doping $\nu$. If $\gamma=
0$ and $\nu$ surpasses a certain minimal value, there are
two critical values of the current, $J_1$ and $J_2$, such
that a wave front is pinned if $J_1\leq J\leq J_2$ and it
moves at a constant velocity $c(J,\nu)$ otherwise. The
velocity $c$ is positive if $J<J_1$ and negative if $J>J_2$.
Furthermore, near the critical currents, $|c|\propto |J-J_c
|^{1/2}$ ($J_c$ is either $J_1$ or $J_2$) \cite{car01}. How
does the disorder correct this picture? The effect of
disorder is to add a term $\gamma D(E_j)\xi_{j+1} - \gamma
[v(E_j)+D(E_j)]\xi_{j}$ to the total current density $J$ in
the dimensionless discrete Amp\`ere equation. Then the
theory in Ref.~\onlinecite{car01} yields an equation similar
to Eq.\ (\ref{cn}) for the front velocity:
\begin{eqnarray}
|c_{R}| = {1\over\pi}\,\sqrt{\alpha \beta (J-J_c) - \gamma
\beta \sum_{n=-L}^M U^{\dag}_{R+n} W_{R+n} },  \label{csr}\\
W_{R+n} = (v_{R+n} + D_{R+n})\, \xi_{R+n} - D_{R+n}
\xi_{R+n+1}.\label{R}
\end{eqnarray} 
Here: (i) $U^{\dag}_{R+n}$ is the left eigenvector
corresponding to the zero eigenvalue of the linearized
equations about the stationary solution at $J=J_c$ (chosen
so that $\sum_{n= -L}^M U^{\dag}_{R+n} U_{R+n} =1$;
$U_{R+n}$ is the right eigenvector), (ii) we define $v_{R+n}
= v(E_{R+n})$, etc., and (iii) $\alpha$ and $\beta$ are
given by Eqs.\ (20) and (21) in Ref.~\onlinecite{car01}. The
values $E_{R+n}$ are those of the stationary electric field
profile at $J=J_c$. The noise term in Eq.\ (\ref{R}) can be
written as $\sum [U^{\dag}_{R+n} (v_{R+n} + D_{R+n}) -
U^{\dag}_{R+n-1} D_{R+n-1}]\xi_{R+n}$, provided we take
$U^{\dag}_{R+n}=0$ for $n<-L$ and $n>M+1$. The noise term
has zero average and a correlation $2\sigma^2/3$, with
$\sigma^2 = \sum [U^{\dag}_{ R+n} (v_{R+n} + D_{R+n}) -
U^{\dag}_{R+n-1} D_{R+n-1}]^2$. Using the previously
mentioned argument of splitting the arithmetic mean in
groups of independent random variables, we can show that
Eqs.\ (\ref{c1}) and (\ref{c2}) hold provided $\sigma$ in
Eq.\ (\ref{c2}) is given by the previous formula and
$(F-F_c)$ is replaced by $(J-J_c)$. Comparisons of the
resulting average front velocity $\overline{c}$ with the
results of numerically solving the SL model are shown in
Figures \ref{fig2} and \ref{fig3} for currents close to
$J_1$ and $J_2$, respectively. 

In conclusion, we have shown that weak disorder changes
qualitatively the wave front depinning transition in
overdamped one-dimensional discrete models. Disorder shrinks
the pinning interval and it changes the critical exponent
for the velocity from 1/2 to 3/2. Whether these features
are robust and hold for strong disorder remains to be seen.
An interesting indication is that the critical exponent 3/2
is obtained independently of the noise strength in mean
field  models of sliding CDW \cite{fis83}. 

 This work was supported by the Spanish DGES grant
PB98-0142-C04-01 and by the Third Regional Research
Program of the Autonomous Region of Madrid (Strategic
Groups Action).

\newpage

\end{multicols}
\end{document}